\documentclass[%
 twocolumn,aps,citeautoscript,superscriptaddress,
 amsmath,amssymb,
 showpacs,floatfix]{revtex4-1}

\usepackage{graphicx}% Include figure files
\usepackage{dcolumn}% Align table columns on decimal point
\usepackage{bm}% bold math
\usepackage[T1]{fontenc}
\usepackage{mathptmx}
\usepackage{etoolbox}

\usepackage[colorlinks,linkcolor=magenta,citecolor=magenta,hyperindex,CJKbookmarks]{hyperref}
\usepackage{ltablex} % for long table
\usepackage{array} % extra row height in table -App D

\usepackage{soul} % to strike out

\newcommand{\bfn} {Ba(Fe$_{1/2}$Nb$_{1/2}$)O$_3$}
\newcommand{\pfn} {Pb(Fe$_{1/2}$Nb$_{1/2}$)O$_3$}
\newcommand{\perovs} {ABO$_3$}
\makeatletter
\begin{document}

\title[Revised crystal structure and electronic properties of BFN ceramics]{Revised crystal structure and electronic properties of high dielectric \texorpdfstring{\bfn}{bfn} ceramics}

\author{Rajyavardhan Ray}
 %\altaffiliation[Also at ]{Dresden Center for Computational Materials Science (DCMS), TU Dresden, 01062 Dresden, Germany.}%Lines break automatically or can be forced with \\
 \affiliation{Leibniz IFW Dresden, Helmholtzstr. 20, 01069 Dresden, Germany.}%
 \affiliation{Dresden Center for Computational Materials Science (DCMS), TU Dresden, 01062 Dresden, Germany.}%
 \email{r.ray@ifw-dresden.de}
 
\author{A. K. Himanshu}%
 \affiliation{Variable Energy Cyclotron Center (VECC), DAE, 1/AF Bidhannagar, Kolkata, India 700064. %\\This line break forced with \textbackslash\textbackslash
}%
 \affiliation{Homi Bhabha National Institute, Mumbai, India 400094.}
 \email{akhimanshu@gmail.com}
 
\author{Golak K. Mandal}
 \affiliation{Department of Physics, T.M. Bhagalpur University, Bhagalpur, Bihar, India 812007}%
 
\author{Uday Kumar}
 \affiliation{Department of Physics, NIT Jamshedpur, Jamshedpur, Jharkhand, India 831014}%

\author{S. N. Jha}
 \affiliation{Raja Ramanna Center for Advanced Technology (RRCAT), Indore, India 452013}%

\author{N. Patra}
 \affiliation{Bhabha Atomic Research Center (BARC), Trombay, Mumbai, India 400085}%

\author{D. Bhattacharya}
 \affiliation{Bhabha Atomic Research Center (BARC), Trombay, Mumbai, India 400085}%

\author{A. B. Shinde}
 \affiliation{Solid State Physics Division, Bhabha Atomic Research Center (BARC), Trombay, Mumbai, India 400085}%

\author{Manuel Richter}
 \affiliation{Leibniz IFW Dresden, Helmholtzstr. 20, 01069 Dresden, Germany.}
 \affiliation{Dresden Center for Computational Materials Science (DCMS), TU Dresden, 01062 Dresden, Germany.}%Lines break automatically or can be forced with \\

\author{P. S. R. Krishna}
 \affiliation{Solid State Physics Division, Bhabha Atomic Research Center (BARC), Trombay, Mumbai, India 400085}%

\date{\today}% It is always \today, today,

\begin{abstract}
{\bfn} ceramics are considered to be promising for technological applications owing to their high
    dielectric constant over a wide range of temperatures.
    However, there exists considerable discrepancy over the structural
    details. We address this discrepancy through a combined x-ray
    diffraction at room temperature and neutron powder diffraction
    measurements in the range from 5K up to room temperature, supplemented by 
    a comparative analysis of the earlier reported structures.
    Our study reveals a cubic structure with space group $Pm\bar{3}m$ at all
    measured temperatures. Further, the x-ray near edge structure and extended
    x-ray absorption fine structure studies on the local environment of the Fe ions is
    consistent with the cubic symmetry. %Our findings highlight the
    %subtle aspects of structural characterization in high pseudosymmetry
    %crystals such as BFN. 
    An appropriate value of $U$ for DFT+$U$ calculations is obtained by comparison with x-ray absorption spectroscopy, which agrees well with the earlier
    reported electronic properties.
\end{abstract}



\maketitle

\section{\label{sec:intro}Introduction}

Members of the perovskite family are known to exhibit a wide variety of
intriguing and unusual physical phenomena, thus, finding applications as
multiferroics \cite{cheong2007,ramesh2007}, catalytic materials
\cite{labhasetwar2015}, energy storage materials
\cite{kostopoulou2018,yang2019} optoelectronic and photovoltaic
materials \cite{snaith2018,roknuzzaman2019}, and dielectric
resonators and filters \cite{cava2001,reaney2006}. With the general
formula {\perovs} (typically, A
= alkali or alkaline earth or lanthanide cations \& B = transition metal
cations), simple perovskites form a 3D network of corner sharing BO$_6$
octahedra while the A-site cations occupy the interstitial 12-fold
coordinated sites between the octahedra. Interestingly, most notable
perovskites which are of high technological interest are not simple
perovskites but rather complex oxides with two different kinds of B-site
cations \cite{davies2008}. Correspondingly, they can be represented with the formula
A(B$_{x}$B$'_{1-x}$)O$_3$.

Complex perovskites possessing high dielectric constant over a
reasonably large temperature range are of particular interest due to
applications as capacitive components, dielectric resonators and
filters. Among the earliest known ferroelectric materials is lead
magnesium niobate, Pb(Mg$_{1/3}$Nb$_{2/3}$)O$_3$ (PMN)
\cite{smolenskii1960}, which belongs to the 1:2 family
($x = 1/3$). It exhibits a classic dielectric relaxation and was
consequently designated as a relaxor ferroelectric. Since then, this
\cite{bokov2006,fu2012,cabral2018} and related
%systems\cite{chung2004,chung2008,singh2012} have been investigated extensively.
systems have been investigated extensively.
Another related compound belonging to the 1:1 family ($x = 1/2$) with Fe
and  Nb transition metal (TM) cations in the B-site,
{\pfn} (PFN)
has also attracted considerable attention \cite{blinc2007,lente2008,bochenek2009}. However, in all these
compounds, lead is an important component. Due to formation of lead
vacancies resulting from evaporation of lead compounds during synthesis
at high temperatures, investigations of stoichiometric compounds are
quite challenging. At the same time, lead compounds have harmful impact
on humans and environment. Therefore, search for lead-free ceramics with
high dielectric constant is of topical interest.

In this article, we focus on barium iron niobate {\bfn} (BFN)
ceramics. BFN based electroceramics are found to exhibit attractive
dielectric and electrical properties over a wide temperature range
\cite{smolenskii1960,tezuka2000,yokosuka1995,saha2001,rama2004,wang2007,eitssayeam2009,eitssayeam2006,ke2013,ganguly2011,raevski2003,demirbilek2004,charoenthai2008,raevski2009,bhagat2010,ke2008,bochenek2009_bfn}. However, there exist contradicting reports with regard to
its structural properties. The structural symmetry of BFN has been
reported to be monoclinic
\cite{saha2001,demirbilek2004,chung2004,ke2008,chung2005,bochenek2009_bfn,chung2008}, cubic
$Pm\bar{3}m$ \cite{yokosuka1995,ganguly2011,wang2007,raevski2009,intatha2010, Khopkar2020}, as well as
face centered cubic $Fm\bar{3}m$ \cite{rama2004,wang2007,ke2013,eitssayeam2006,eitssayeam2009} 
at room temperature. Furthermore, any correlation between the synthesis method and the reported structural
symmetry is unavailable.

We address this discrepancy over the
structure by carrying out a detailed overview of the reported crystal structures and
its possible dependence on the choice of synthesis method and
parameters. We argue that the source of such a discrepancy is likely
due to possible pseudosymmetry of BFN as observed earlier for
many other perovskites, and can be resolved
through a careful analysis of the X-ray
diffraction (XRD) and neutron powder
diffraction (NPD) measurements. Subsequently, we obtain
the space group symmetry of BFN by carrying out combined XRD and NPD measurements. We find that BFN
crystallizes in the cubic $Pm\bar{3}m$ space group (No. 221) at ambient
temperature and that the chemical formula matches with the single
perovskite Ba(Fe$_{1/2}$Nb$_{1/2}$)O$_3$, disagreeing with the earlier reports of
monoclinic and cubic $Fm\bar{3}m$ structures. This further implies that the
occupation of the B-site by the TM ions Fe and Nb is
random. Further details regarding the local structure are obtained
using extended x-ray absorption fine
structure (EXAFS) and x-ray near edge structure (XANES) spectroscopy.

A detailed investigation of the electronic and optical properties is
also carried out using a combination of diffuse reflectance spectroscopy
in the UV-Vis-NIR range and X-ray
absorption spectroscopy (XAS) together with Density Functional Theory
(DFT). The electronic properties are found to be consistent with earlier
reports. In order
to properly account for the electron-electron correlation effects within
DFT, the GGA$+U$ functional is employed where the appropriate value of
$U$ is obtained by comparison with the UV-Vis-NIR and O-1$s$ XAS spectra.

Our study points out the existing knowledge gaps and highlights the
subtleties regarding the structural characterization of BFN and high
pseudosymmetric complex perovskites in general. The XRD pattern alone may not be
sufficient to correctly characterize the structural symmetry. A combined XRD and NPD
measurement for BFN strongly suggest that BFN crystallizes in a cubic
structure with $Pm\bar{3}m$ space group. 
The local structure, as obtained by EXAFS, points to
disordered BO$_6$ octahedra where the distortions are non-polar.

This article is organized as follows: In Sec \ref{sec:methods_expt}, we present the
synthesis and experimental methods, which is followed by details of the
DFT calculations employed in this study in Sec \ref{sec:methods_dft}. In
Sec \ref{sec:results}, we
present our results and discuss their implications, and finally conclude
in Sec \ref{sec:concl}.

\section{\label{sec:methods_expt}Synthesis \& Experimental Methods}

BFN ceramics were synthesized by
the columbite precursor method. The precursor FeNbO$_4$ was
synthesized by solid state reaction of reagent grade Fe$_2$O$_3$
(99.999\%) and Nb$_2$O$_5$ (99.99\%) by mixing them in predetermined
amounts. The mixture was wet milled in acetone for
a day, and calcined at 1150 $^{\circ}C$ for 5 hr. In the second stage,
FeNbO$_4$ and BaCO$_3$ were mixed in stoichiometric ratio and calcined
at 1200 $^{\circ}C$ for 8 hr for formation of BFN. The pellets of the
calcined material were finally sintered at 1250 $^{\circ}C$ for 4 hr.
After each stage of calcination and final stage sintering, the phases were
confirmed with XRD taken at room temperature on Rigaku Miniflex II. 
The data were obtained in the range of $20^{\circ} \le 2\theta \le 80^{\circ}$, with a step size of
$0.02^{\circ}$, using
a Cu K$\alpha$ source (average $\lambda = 1.5418$ {\AA}) with a beam current of 15 mA and at a potential of
30 kV.

The sintered pellets were crushed
to form powder for the Neutron Powder diffraction (NPD) measurement. NPD
was performed on
PD-2 diffractometer \cite{krishna2002} at the Dhruva reactor, Bhabha Atomic Research
Center (BARC), India, using neutrons of wavelength 1.244 {\AA} in a
temperature range of 300 K to 5 K.  Rietveld refinement
of the XRD \& NPD data was carried out with the help of
the FullProf program \cite{fullprof}. The background was fitted with 6-coefficients
polynomials function, while the peak shapes were described by
pseudo-Voigt profiles. In all the refinements, scale factor, lattice
parameters, and thermal parameters were
varied. Positional parameters were kept fixed since all atoms occupy
special positions. In the initial stage, occupancy parameters of all the ions were kept fixed during
refinement. It was possible to refine all the
parameters together. In the next step, occupancy parameters were varied
to ascertain the quality of the fit. No significant deviation were
found, implying the robustness of the obtained structural model. 
Other cubic structural model suggested in earlier works were also considered which, however, 
led to slightly poorer fit. 

The bandgap of the room temperature phase
was determined from the diffuse reflectance
measurements in the UV-Vis-NIR range as suggested by Davis and Mott
\cite{Davis1970}.
The diffuse reflectance spectrum was obtained using a Perkin-Elmer 950
UV/Vis/NIR spectrophotometer. It was later converted to an equivalent
Kubelka-Munk (KM) absorption spectrum \cite{ray2017,barton1999}:
\begin{equation}
F(R_{\infty}) = A \frac{(h\nu -  E_g)^n}{h\nu} \,\,,    
    \label{eqn:km}
\end{equation}
where, $F$ is the Kubelka-Munk function of the relative reflectance of
the sample with respect to a reference $R_{\infty}=R_{\rm
sample}/R_{\rm reference}$, $A$ is a proportionality constant, $h\nu$ is
the incident photon energy and $E_{\rm g}$ is the bandgap. The values of
$n$ can be appropriately chosen to be $n =1/2$ or $2$ corresponding,
respectively, to direct or indirect allowed transitions. Thus, the bandgap can
be obtained by inverting the above relation (Eq. (\ref{eqn:km})) and, plotting
$[F(R)h\nu]^{1/n}$ with suitable $n$ as a function of energy. The
intercept of the linear part of the spectrum near the absorption edge on
the energy axis provides an estimate of the optical gap
\cite{Davis1970,ray2017,barton1999}.

To further clarify the local structure of BFN, the X-ray absorption spectroscopy
(XAS) techniques were used. The Fe
L$_{2,3}$-edge and O K-edge X-ray Absorption Near Edge
Structure (XANES) spectra were recorded in the total electron
yield (TEY) mode at the soft X-ray absorption spectroscopy beamline
(BL-01) of the Indus-2 at Raja Ramanna Center for Advanced Technology
(RRCAT), Indore, India. The synchrotron based XANES and Extended X-ray Absorption Fine Structure (EXAFS)
studies at the Fe K-edge ($\sim 7112$ eV) of the {\bfn} samples were carried
out in fluorescence mode at the Energy Scanning EXAFS beamline (BL-09)
at Indus-2 Synchrotron source (2.5 GeV, 100 mA) at RRCAT, which operates
within the photon energy range 4-25 keV \cite{basu2014,poswal2014}.  The beamline optics consists of
a Rh/Pt coated meridional cylindrical mirror used for the collimation of
the beam. The collimated beam is subsequently monochromatized by a Si
(111) ($2d = 6.2709$ {\AA}) based double crystal monochromator (DCM). The
pre-mirror used prior to the DCM also helps in higher harmonic
rejection. Second crystal of the DCM is a sagittal cylindrical crystal,
which is used for horizontal focusing of the beam while another Rh/Pt
coated bendable post mirror facing down is used for vertical focusing of
the beam at the sample position. For the measurement in the fluorescence
mode the sample was placed at $45^{\circ}$ to the incident beam. The fluorescence
signals ($I_f$) were collected using a Si drift detector placed $90^{\circ}$ to the
incident beam ($45^{\circ}$ to the surface of the sample) and the 1st ionization
chamber measures the incident beam ($I_0$). In this way the X-ray absorption
coefficient of the sample is determined by $\mu = I_f/I_0$, and the spectrum was
obtained as a function of energy by scanning the monochromator over the
specified range.

In order to get the oscillation in the
EXAFS spectrum, the energy dependent normalized absorption
coefficient $\mu(E)$ was converted to energy dependent absorption
function $\chi(E)$ and then to the wave number dependent absorption
coefficient $\chi(k)$. Finally, $k^2$ weighted  $\chi(k)$ spectra were Fourier transformed in real space to generate
the $\chi(E)$ versus $R$ (or FT-EXAFS) plots in terms of the real distance
from the center of the absorbing atoms. The analysis of the EXAFS data
was performed following the standard procedure \cite{konigsberger1988,
kelly2008} using the
available IFEFFIT software package \cite{newville1995} 
The generation of the theoretical EXAFS spectra 
was done using the FEFF 6.0 code. During the fitting process the EXAFS parameters
{\it i.e.} bond distance ($R$) between the respective atomic pairs along a
particular scattering path and the disorder factor (Debye-Waller factor
$\sigma^2$)  
were varied independently. The FT-$k^2\chi(k)$ spectra have been
fitted using the standard model obtained from the Neutron diffraction
study. For the purpose of fitting, data in the $k^2$ weighted
$\chi(k)$
spectra have been taken within the $k$ range 2.5 - 8.5 {\AA}$^{-1}$ with the phase
uncorrected $R$ range 1 - 3.7 {\AA}. The goodness of the fit in the
above process is 
generally expressed by the $R_{\rm factor}$ which is defined as:
\begin{equation}
    R_{\rm factor} = \sum \frac{[{\rm Im} (\Delta \chi(r_i) )]^2 + [{\rm Re} (\Delta \chi(r_i))]^2}{[{\rm Im} \chi_{\rm dat}(r_i)]^2 + [{\rm Re} \chi_{\rm dat}(r_i)]^2}
\end{equation}
where $\Delta \chi = \chi_{\rm dat} - \chi_{\rm th}$ is the difference
between the experimental and theoretical $\chi(R)$ values and Im and Re
refer to the imaginary and real parts of the respective quantities.

\section{\label{sec:methods_dft}Computational Details} 

The obtained structure after the refinement of the XRD and NPD data is
the perovskite $Pm\bar{3}m$ structure, which implies that the B-site cations,
Fe and Nb, share the same Wyckoff position ($2a$) and are randomly
occupied. Within DFT, description of ``disordered" systems remains a
challenge. A theoretical analysis of such a structure requires
calculations with large supercells and for different possible ensembles
for B-site occupancy \cite{Morrow2020}. We consider $2 \times 2 \times 2$ supercell structures with 1:1
occupation of the B-site by Fe and Nb atoms. The relative energy difference between these structures is
found to be $\lesssim 0.25$ eV/f.u. (see Table \ref{table:supercell_str} for details).
Due to limited size of the
supercell and cubic symmetry of the compound, each of the resulting
structures possesses ordering of the Fe and Nb ions along different
directions.  
Nevertheless, insights into the
electronic properties of BFN can be obtained by considering the lowest
energy structure among these, which is the adopted strategy here. The
lowest energy configuration corresponds to an ordered arrangement of Fe
and Nb ions in the Wyckoff position $2a$ along all the crystal directions,
leading to a face-centered cubic $Fm\bar{3}m$ structural model. We focus on
this structure model for detailed investigations of the electronic properties. 

It is important to note that, within the adopted computational scheme,
although the lattice parameters are doubled (unit cell is doubled), the
position of the O atoms are kept fixed at ($1/4$, $0$, $0$) corresponding to
the ideal face-centered cubic structure since there are no free
parameters in the atomic positions of the $Pm\bar{3}m$ structure.
The optimal volume obtained in the Generalized Gradient Approximation
(GGA) for the $Fm\bar{3}m$ phase is approximately 4.3\%
larger than obtained from XRD and NPD refinements (see Fig.
\ref{fig:vol-opt}), which is a typical difference. In the
following, however, we consider only the experimental lattice constant.

All calculations were carried out within the Perdew-Burke-Ernzerhof
(PBE)  implementation \cite{pbe96} of GGA using the Full-potential
Augmented Plane Waves plus local orbital (FP APW+lo)
method as implemented in WIEN2k 17.1 \cite{wien2k}. A $12 \times 12
\times 12$ $k$-mesh of the full
Brillouin zone was used for the numerical integration together with 
$R_{\rm MT} \times K_{\rm Max}  = 7.0$. An energy cut-off of -6.0 Ryd was chosen to separate the
core and valence states. Electron-electron correlation effects were
included via the spherically symmetric self-interaction correction
(SIC) \cite{anisimov1993} double counting term of the GGA$+U$ functional.
The self-consistency is better than 0.0001 e/a.u.$^3$ for the charge density and the stability is
better than 0.01 mRy for the total energy per unit cell.
For comparison with the O K-edge XAS spectra, a constrained
calculation was also carried out with a core-hole in only one of the O-$1s$
states in the unit cell.

\section{Results \& Discussions}
\label{sec:results}

\subsection{\label{sec:str}Structural Properties}

We begin with an overview of the reported crystal structures of BFN
ceramics. The most commonly employed synthesis technique is the solid-state
reaction method, which 
may \cite{ke2008,ke2009,chung2004,chung2008,ke2010,liao2010} or may 
not
\cite{yokosuka1995,saha2001,rama2004,wang2007,eitssayeam2006,eitssayeam2009,intatha2010,ganguly2011,
Khopkar2020} 
involve the use of a B-site precursor.  Additionally,
conventional ceramic method \cite{bhagat2010, tezuka2000}, powder
calcination\cite{bochenek2009}, sol-gel method \cite{chung2005}, chemical
synthesis \cite{charoenthai2008}, mechanical
triggering \cite{bochenek2015}, and mechanochemical \cite{bochenek2018} routes have also been employed.
The reported structural
data comprises of perovskite cubic $Pm\bar{3}m$ structure, double
perovskite cubic  $Fm\bar{3}m$ as well as monoclinic structures.
However, any correlation
between the choice of synthesis method or synthesis parameters and the resulting structure
could not be established. For brevity, we tabulate the synthesis method,
synthesis parameters and the resulting crystal structures in Table
\ref{table:review} (Appendix \ref{sec:AppE}).

Even within the standard solid-state reaction techniques (without the
involvement of a precursor), both cubic and 
monoclinic structures have been reported. The earliest work by
Yokosuka in 1995 reported a cubic perovskite $Pm\bar{3}m$
structure \cite{yokosuka1995}. However, in 2002, Saha and Sinha
reported a monoclinic structure \cite{saha2001}. The two methods
differ in values of calcination and sintering temperatures used
during the synthesis. However, subsequent works
using similar calcination temperatures as used by Saha and Sinha
have reported either cubic $Fm\bar{3}m$ \cite{rama2004,eitssayeam2006},
 or monoclinic \cite{demirbilek2004,bochenek2009_bfn} or cubic $Pm\bar{3}m$
structures \cite{raevski2003,eitssayeam2009,bhagat2010}.
With regard to the calcination temperature, it is found that
temperature above 900$^{\circ}$ is required for the formation of BFN
samples \cite{eitssayeam2009}. However, this doesn't influence the
resulting structural symmetry and all the reported studies meet 
this requirement. Similarly, the
sintering temperature influences only the quality of the resulting
sample, it has no effect on the crystal structure \cite{ke2008}. 

The use of a B-site precursor has been proposed as an effective
method to obtain high quality samples in Pb-based perovskites, for
example the Pb-counterpart,
{\pfn} (PFN) \cite{raymond2003}. In both PFN and BFN, the columbite
precursor FeNbO$_4$ is used, which 
crystallizes in a monoclinic (wolframite $P2/c$ or AlNbO$_4$-type
$C2/m$), orthorhombic (ixiolite $Pbcn$), or tetragonal (rutile
$P4_2/mnm$) structure. Among these, the use of monoclinic wolframite
structure and the orthorhombic rutile structures have received some
attention. For BFN, a comparison between the samples synthesised using the
columbite precursor and direct mixing of initial compounds (oxides
and carbonates) suggests that columbite precursor methods should be
the method of choice for technological applications since it leads to low dielectric
loss \cite{ke2008}.

During the synthesis of BFN using the columbite precursor, Ke {\it
et al.} \cite{ke2008,ke2009,ke2010} have carefully chosen the monoclinic
precursor (calcination temperature of 1075$^{\circ}$ for the precursors) 
as if the structural symmetry of the
precursor and the resulting BFN samples are correlated. While they find
the resulting BFN ceramics possessing monoclinic structure, this
causal relationship has never been established. In fact, for the related PFN
ceramics, it was explicitly shown that the structural symmetry of the precursor does
not play any role in the structure of resulting
ceramics \cite{raymond2003}.

Other works which use similar calcination temperature for
the precursor have found either a monoclinic \cite{chung2008} or a cubic
$Pm\bar{3}m$ structure \cite{liao2010}. The difference in the structural
symmetry could be due to different choice of calcination temperature
during the second-stage mixing of the precursor with the BaCO$_3$. The higher
calcination temperature during this stage led to a $Pm\bar{3}m$
structure. However, this apparent dependence of the crystal structure of
BFN on the crystal structure of the precursor is in sharp
contrast to PFN, and possibly misleading.

A closer inspection of the available XRD patterns for different reported
crystal structures irrespective of the synthesis route, reveals that the source of such discrepancy could
be related to high pseudosymmetry of the perovskite BFN ceramics. The XRD patterns
presented in Ref. \cite{yokosuka1995} for $Pm\bar{3}m$, Ref.
\cite{saha2001} for monoclinic, and Ref.
\cite{tezuka2000} for $Fm\bar{3}m$ structures are almost identical.
In fact, such conflicting structural characterization is of concern and has remained a challenge
for a long time \cite{barnes2006}; one of the notable example in recent times being 
the Van Vleck paramagnet Sr$_2$YIrO$_6$ \cite{cao2014,corredor2017}. 

For compounds with possible high pseudosymmetry, the characterization
should be based on a careful analysis of the presence of 
additional reflection peaks and/or peak splittings. It is interesting to note
that among all the claims for monoclinic structure, such an evidence
for a monoclinic structure is available only in  Ref.
\cite{raevski2009b}, where the XRD data shows clear peak splitting
of the (220) and (222) reflection peaks. However, such splitting of the
(220) reflection peak is absent in other claims. Moreover, some of the 
authors in Ref. \cite{raevski2009b} have also claimed cubic structures based on similar
synthesis parameters. Arguably, the realization of the
monoclinic structure is possibly due to some experimental
factors other than the calcination and sintering temperatures used. We,
therefore, consider this as an outlier. (This data has also been omitted
from Table \ref{table:review}.)

%\begin{widetext}

\begin{figure*}[t!]
    \centering
    \includegraphics[angle=0,width=0.805\textwidth]{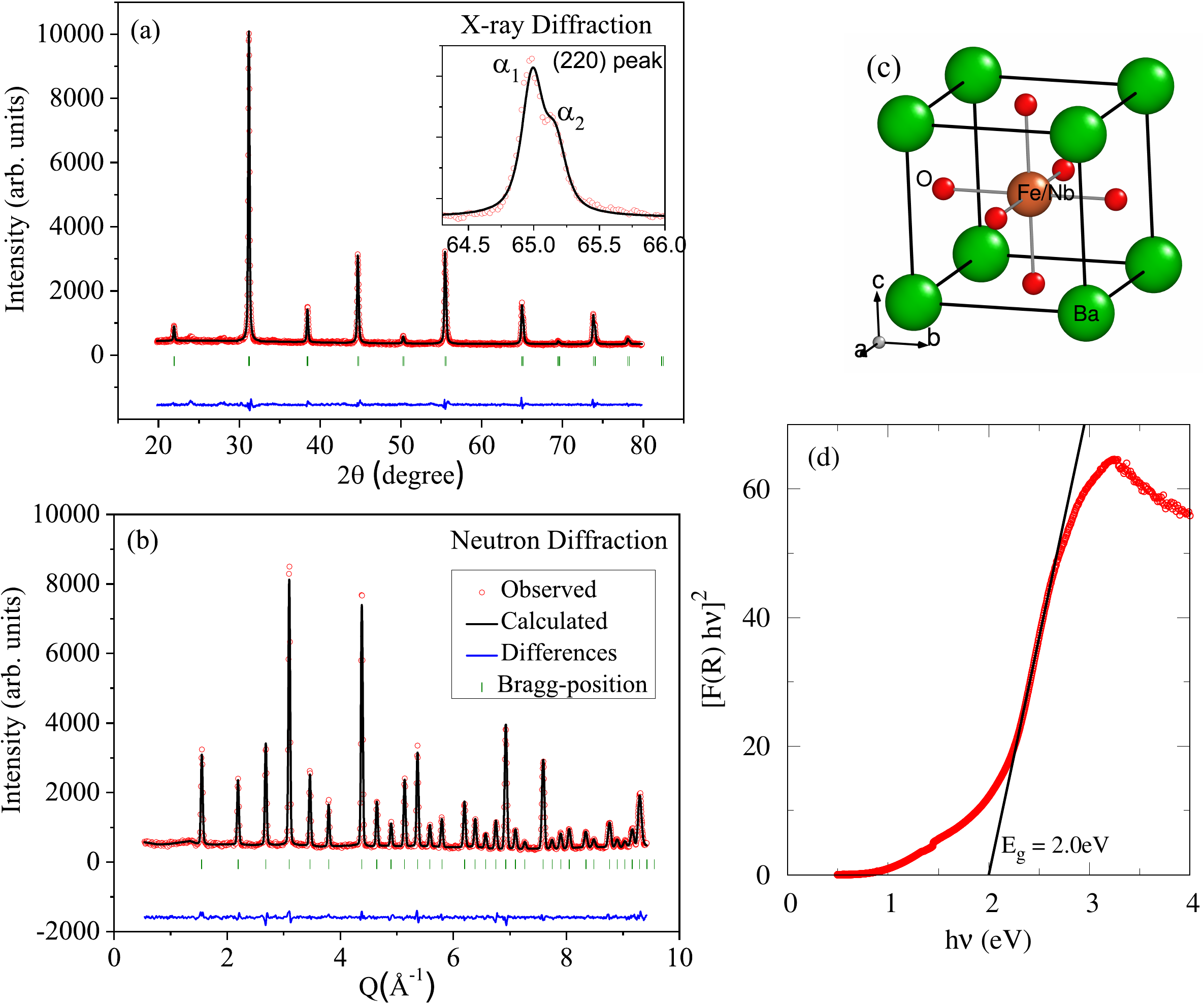}
    \caption{ Room temperature (a) XRD data, (b) NPD Data and (c) the resulting crystal structure of BFN. 
        The inset in (a) shows the non-split ($220$) peak confirming the cubic structure. The
        doublet $\alpha_1$ and $\alpha_2$ corresponds to the Cu K$\alpha_1$ and K$\alpha_2$
        beams. (d) Kubelka-Munk function vs energy showing a direct bandgap of approximately 2eV.
    }
   	\label{fig:npd}
\end{figure*}
\begin{table*}[ht!]
    %\tiny
    \caption{Structural parameters obtained from NPD measurements. 
    }
\label{table:str_npd}
\small
%\begin{tabular}{lccc}
%\begin{tabular*}{0.85\textwidth}{lccc}
\begin{tabular*}{0.780\textwidth}{p{4.3cm} p{3.4cm} p{3.4cm} p{4.0cm} }
 \hline 
    \hline
    \multicolumn{4}{l}{\bf A. Refinement parameters} \\
 \hline \hline
    {\bf Param} & {\bf 300 K (ND)} &  {\bf 5 K (ND)} & \\
    \hline
    Spage group  & $Pm\bar{3}m$ (No. 221) & $Pm\bar{3}m$ (No. 221) &  \\
    $a = b = c$ ({\AA}) & 4.0556(2) & 4.0466(1) & \\
    $\alpha = \beta = \gamma$ & 90$^{\circ}$ & 90$^{\circ}$ & \\
    Final R-indices: & &  & \\
    R$_{\rm p}$ & 3.94 & 3.92   & \\
    R$_{\rm wp}$ & 5.02 & 5.10   & \\
    R$_{\rm exp}$ & 3.69 & 3.68   & \\
    Goodness of Fit (GoF) & 1.85 & 1.91  & \\
 \hline \hline
    \multicolumn{4}{l}{\bf B. Structural parameters at 300 K from
    neutron diffraction} \\
 \hline \hline
   
    {\bf Atom} & {\bf Position}  & {\bf Occupancy} & {\bf $B_{\rm iso}$ 300K \& 5K} \\
      \hline
    Ba & (0,0,0) & 1.0 & 0.58(4) \& 0.22(3) \\
    Fe & (1/2,1/2,1/2) & 0.50 & 0.59(4) \& 0.35(3)  \\
    Nb & (1/2,1/2,1/2) & 0.50 & 0.59(4) \& 0.35(3)  \\
    O & (1/2,1/2,0) & 1.0 & 0.68(3) \& 0.39(3) \\
    \hline
    \hline
\end{tabular*}
\end{table*}
%\end{widetext}

Empirically, tolerance factor, defined as:
\begin{equation}
    t = \frac{r_{\rm A} + r_{\rm O}}{\sqrt{2} (\bar{r}_{\rm B,B'} +
    r_{\rm O})},
\end{equation}
can be a useful guide to ascertain if the resulting structure could be
cubic \cite{ray2017,bartel2019}. Here, $r_{\rm A}$, $r_{\rm O}$, and $\bar{r}_{\rm B,B'}$ are
the ionic radii of the A-site cation, O ion and average of the
B-site cations, respectively. $t=1$ represents the ideal case of a cubic unit cell with perfect matching between
    the A- and B-site ionic radii. Therefore, if $t \approx 1$, a cubic structure is
expected. For $t < 1$, the size mismatch between the radii of the A- and B-sites leads to octahedral
distortions, lowering the symmetry to orthorhombic, tetragonal or monoclinic structures. If $t> 1$, the A-site ion is too big (compared to the B-site cations), leading to
distortion of the structure towards a hexagonal or tetragonal symmetry. Note, however, that the
accuracy of this method is approximately $83\%$ for perovskite oxides.\cite{bartel2019} For BFN, it turns out that $t = 1.04$, suggesting that a
cubic structure is possible. 
The degree of cation ordering is determined by the ratio of ionic radii
of the B-site cations, large difference usually leads to long range
cation ordering.
By the fact that both
the B-site cations Fe$^{3+}$ and Nb$^{5+}$ have nearly the same
ionic radii, a large degree of cation disordering is
expected.

A possible way to clarify the structural details (especially cation ordering
and atomic parameters) is to perform a careful analysis of simultaneous XRD
and NPD measurements since they provide complementary information. The cubic and
    monoclinic crystal symmetries lead to distinct reflection peaks (\textit{e.g.} splitting of the (220) peak
    only for monoclinic unit cells) and can be distinguished by a high resolution XRD as well as NPD
    data. On the other hand, as NPD is more sensitive to the
positions of O atoms,\cite{barnes2006} the oxygen position can be accurately determined in the NPD
data, which differ in the cubic $Pm{\bar 3}m$ and $Fm{\bar 3}m$ symmetries, as discussed below.
This is precisely the strategy
employed in this work. 

Our choice of the synthesis parameters is same as
in Ref. \cite{ke2008}, except for the calcination temperature
during the preparation of the precursor (see Sec. \ref{sec:methods_expt}). 
We checked that this does not affect the structural symmetry of the
precursor, which was found to be monoclinic (see App. \ref{sec:AppA} for details).

Figures \ref{fig:npd}(a) \& (b), respectively, show the XRD and NPD
patterns for the synthesized BFN samples at room temperature. 
The Rietveld refinement of the XRD and the NPD data reveals that BFN
crystallizes in the cubic $Pm\bar{3}m$ (No. 221) structure at ambient
temperatures. In the XRD data, the (220) peak at $2\theta \sim 65^{\circ}$ is not split [shown in
the inset of Fig. \ref{fig:npd}(a)], confirming
the cubic structure. Similarly, no superlattice reflection peaks corresponding to a monoclinic
structure were observed in the NPD data either. 

The distinction between the cubic $Pm\bar{3}m$ and
$Fm\bar{3}m$ structures
arises from long range cation
ordering in the B-site, leading to different sizes of the BO$_6$ (and B$'$O$_6$)
octahedra. The $Fm\bar{3}m$ structure corresponds not only to a doubled cell but the
    position of the oxygen
    atom ($x$ coordinate) is also a variable (Wyckoff site $24e$). Depending on the ionic radii,
    electronegativity, and hybridization with oxygen for the transition metal cation, 
    the B-O and B$'$-O bonds are of different length. In comparison, in the $Pm\bar{3}m$ structure,
    the oxygen coordinate is fixed at the center of the B/B$'$ positions. As the oxygen atom
scattering has more information content in NPD patterns compared to XRD patterns, it is possible to 
decipher the position of the oxygen atoms, and thus, also the
appropriate structural model. Any deviation of oxygen atom $x$-coordinate in the $Fm\bar{3}m$
structure should lead to non-zero intensities in the super lattice reflections. It should be noted that,
in the investigated samples, both
the structural models models fit the data well (based on the goodness of fit). However, we find
that all the super lattice reflections in
$Fm\bar{3}m$ model fit have zero intensity (see Appendix \ref{sec:AppB}).
Consequently, the
oxygen positions are such that the (average) size of the BO$_6$ and B$'$O$_6$
octahedra are identical despite different electronegativities of
the B and B$'$ cations, thus, leading to the $Pm\bar{3}m$ structure
model. %\oth{[Oxygen occupation? (Add \%-age occupation, if possible)]}
Similarly, monoclinic phase can also ruled out as
any other superlattice reflection other than for perovskite cubic $Pm\bar{3}m$ in NPD and cubic
symmetry in XRD was not observed. Moreover, no signature of any subtle structural transition were
observed based on the analysis of the full-width half-maxima (FWHM) of the Bragg peaks in the NPD
data at 300 K and at 5 K (see Appendix \ref{sec:AppB}).
Therefore, based on a combined XRD and NPD data analysis, we unambiguously establish the structure
of {\bfn} ceramics to be cubic $Pm{\bar 3}m$.

From the NPD data refinement, the lattice constant
was found to be 4.0556 {\AA} at 300 K and the chemical formula matches with that
of a single perovskite Ba(Fe$_{0.5}$Nb$_{0.5}$)O$_3$ as reported in some of the
previous studies. Also, no structural transition was observed down to 5 K. 
The structural details are presented in Table
\ref{table:str_npd} and the  
corresponding crystal structure is shown in Fig. \ref{fig:npd}(c).

\subsection{\label{sec:km}Optical bandgap (UV-Vis)}
Figure. \ref{fig:npd}(d) shows the KM absorption spectrum as a function of the
incident photon energy for $n=1/2$ corresponding to the direct allowed
transition.
The gap is estimated to be approximately 2.0 eV as obtained from the
intercept of the linear part on the energy axis. It is important
to note that, for $n=2$ (indirect allowed transitions), lower estimated gaps could
be obtained. However, they are not in agreement with the dark brown
color of the sample. On the other hand, a gap value of 2.0 eV is
consistent with the sample color. Therefore, in the following, we consider the gap value of
2.0 eV to be the likely bandgap of the synthesized BFN samples. 

\subsection{Electronic Properties}

\subsubsection{\label{sec:xas}XAS and XANES}

\begin{figure*}[ht!]
    \centering
    \includegraphics[angle=0,width=1.000\textwidth]{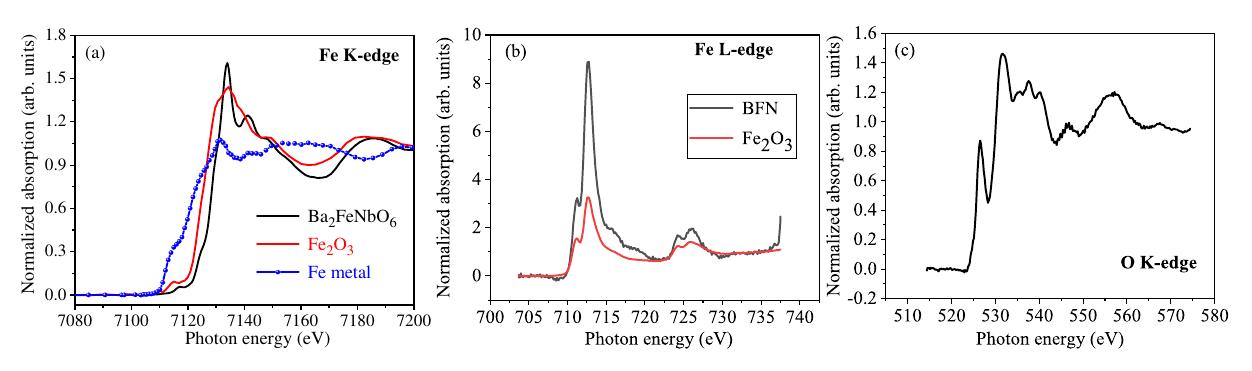}
    \caption{(a) Normalized XANES spectra at Fe K-edge compared to
        reference material Fe$_2$O$_3$ and Fe metal, and (b) XAS for Fe
        L-edge in BFN and Fe$_2$O$_3$. (c) XAS spectrum for the O K-edge
        in BFN. See text for further details.
    }
   	\label{fig:xanes}
\end{figure*}

The valency of constituent atoms can also be analyzed using X-ray
absorption spectroscopy (XAS). XAS has additional advantages since it is
not only elemental selective in analyzing the electronic (valence)
states but also capable of probing the local structures simultaneously
through X-ray absorption near edge structure (XANES) and extended X-ray
absorption near edge structure (EXAFS). Fig. \ref{fig:xanes}(a) shows the XANES
spectra of the BFN sample measured at Fe K-edge along with that of Fe
metal foil and Fe$_2$O$_3$ powder. The absorption edge lies at approximately
$\sim 7128$ eV for BFN and $\sim 7123.5$ eV for the reference material
Fe$_2$O$_3$. The
appearances of features just below the absorption edge is
usually referred to as `white lines' \cite{tranquada1987} and corresponds to electronic
(atomic-like) transitions from the core $1s$ state into a quasi-bound state
with $3d$ character. While such transitions are forbidden within the
electric dipole approximation for an ideal octahedral crystal field
environment, slight distortions in the MO$_6$ (M = Fe, Nb) octahedra may lead
to such pre-edge feature. The pre-edge features at $\sim 7115$ eV are due to
the $1s$ transition into unoccupied O-2$p$ - M-3$d$/4$sp$ hybridized states,
which have $p$-component projected at the sites as observed in many
transition metal oxides. Presence of such hybridized states with mixed
$pd$ character near the
band edges is also seen in the DFT calculations (see Fig. \ref{fig:dft}(a)), as
discussed  in detail later.

After the pre-edge background (7115 eV) subtraction, the spectra are
normalized to unity, averaging the signal between the edge jumps
and about 50 eV above the white line. We also measure the spectra for Fe
metal foil as reference. It is evident from a comparison between
these spectra that the Fe K-edge in the sample lies slightly above that of
Fe$_2$O$_3$, implying that the oxidation state of Fe in the present sample is
$+3$. %This is consistent with lack of evidence for Fe$^{2+}$ impurities %in the XPS data.

The intensity and peak position of the pre-edge depends critically on the local
structure and the oxidation state \cite{shuvaeva2017}. Since the Fe ions
in Fe$_2$O$_3$ are known to be highly non-centrosymmetric, 
the fact that the intensity of the pre-edge for BFN is much weaker than that for
Fe$_2$O$_3$ is indicative of the fact that the off-center displacement
of Fe ions is negligible. This is further substantiated by the fact the 
peak lies at somewhat higher energy than for
Fe$_2$O$_3$, similar to the primary absorption peak, and that the
primary absorption (dominant) peak is much steeper. 
These findings essentially rule out any non-centrosymmetric position
of the Fe$^{3+}$ ions and are consistent with an ideal $Pm\bar{3}m$ structural
characterization \cite{shuvaeva2017}.

For the L-edge shown in Fig. \ref{fig:xanes}(b), the possible transitions from
2$p_{1/2}$  and 2$p_{3/2}$ states to the $d$ states lead to L$_2$ and
L$_3$ peaks in
the XAS spectra. These peaks further split into two peaks which 
is a characteristic at the L$_{2,3}$ edge of all $d^n$ compounds in
an octahedral or tetrahedral environment due to transitions %from 2$p_{3/2}$ 
to $t_{2g}$ and $e_g$ levels. This ligand field splitting is
estimated to be $\sim 1.5$ eV. The spectrum is very similar
to Fe$_2$O$_3$ as would be expected for Fe$^{3+}$ ions. 

Figure \ref{fig:xanes}(c) shows the normalized XAS for the O K-edge of BFN. The
low-energy part of the XAS spectrum is dominated by the energy positions of the
TM-$d$ states due
to the strong covalent bonding among the transition metal TM-$d$ and
O-2$p$ orbitals. The low-energy features lying at $\sim 526.5$ eV and
$\sim 531.6$ eV
should be related to the low energy transition from O-1$s$ state to
the O-2$p$ states hybridized with the $d$-states from the transition
metals Fe and Nb. These features are also confirmed by the first
principles calculations and will be used to obtain a reliable
estimate of the strength of $U$ for GGA$+U$ calculations, as discussed in the
following. The features at energies greater than $\sim 536$ eV can be
attributed to transitions to the higher lying hybridized O-2$p$ $-$
TM-$sp$ states. 
These finding are consistent with the DFT calculations. 

\subsubsection{\label{sec:dft}DFT}

\begin{figure*}[ht!]
\centering
\includegraphics[angle=0,width=0.90\textwidth]{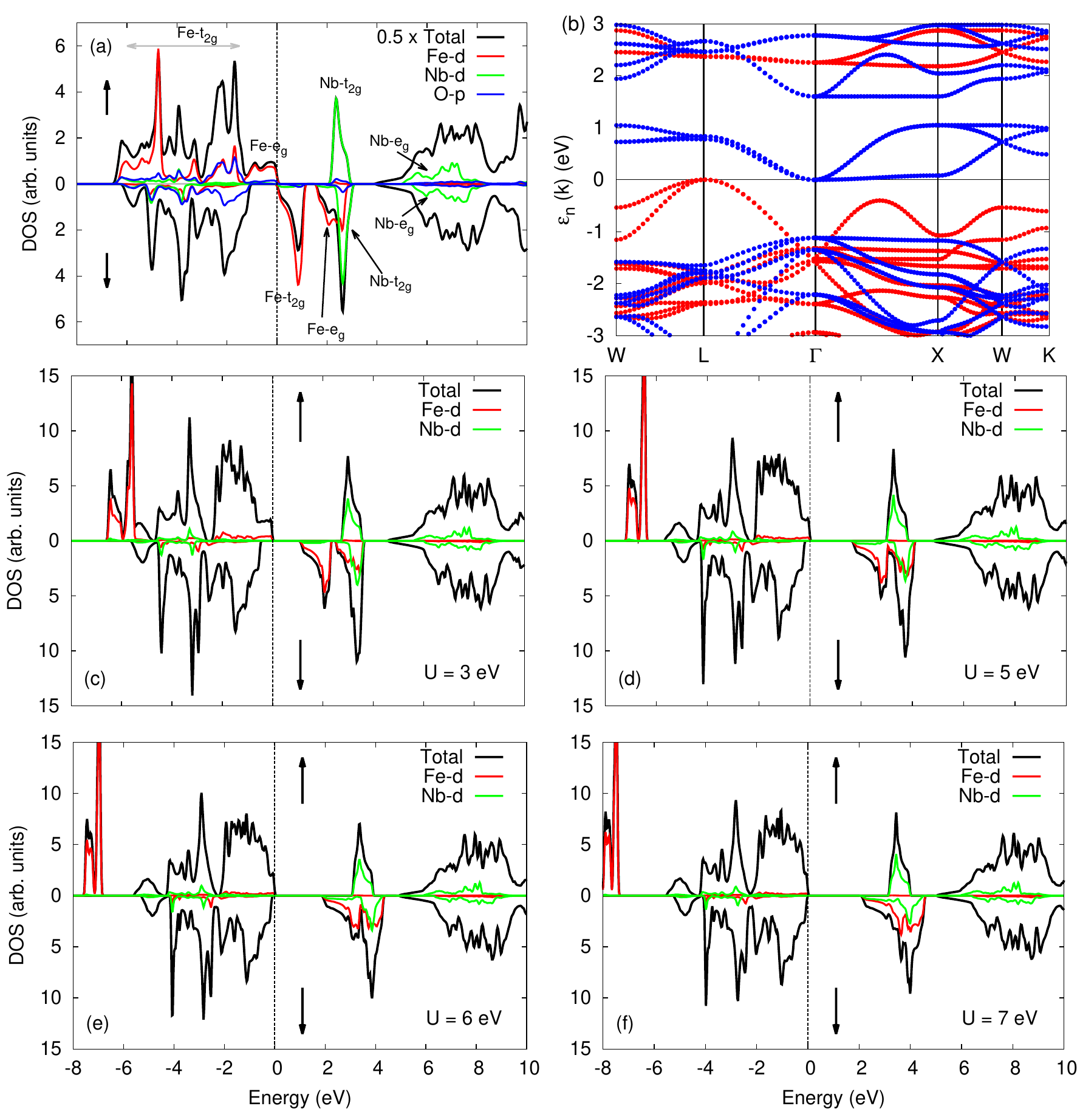}
\caption{The electronic properties of BFN: (a) spin-projected total
    and (atom-resolved) partial density of states (DOS) obtained
    within GGA scalar relativistic calculations ($U=0$). The spin
    components are represented by up and down arrows. The
    corresponding band structure is shown in (b), where spin up
    (down) bands are shown in red (blue). The high-symmetry points
    are located at: W ($\pi/a$,$2\pi/a$,0), L
    ($\pi/a$,$2\pi/a$,$\pi/a$), $\Gamma$ (0,0,0), X (0,$2\pi/a$,0)
    and K ($3\pi/2a$,$3\pi/2a$,0). (c)-(f) The evolution of
    the total and TM-$d$ partial DOS with $U$. The Fermi energy is
    set to $\epsilon_{\rm F}=0$.
}
   	\label{fig:dft}
\end{figure*}

%\end{widetext}

The total and
partial density of states (DOS) within GGA ($U=0$) is shown in Fig.
\ref{fig:dft}. The
states across the Fermi energy are largely composed of Fe-3$d$ states. Due
to the octahedral ligand field exerted by the O atoms, the $d$-shell
splits into $t_{2g}$ and $e_g$ levels, $t_{2g}$ being lower in energy.
For the up-spin channel, the $t_{2g}$ and $e_{g}$ states lie between the
Fermi energy $\epsilon_{\rm F}$ and -6 eV. In the down spin channel, the ligand
field is estimated to be approximately 1.5 eV,
which is in agreement with the corresponding value estimated for XAS
(see Fig. \ref{fig:xanes}(b)). Consistent with the
formal valency of +3, the Fe-$d$ states are half-filled (fully populated
by up-spins) and, therefore, experience large exchange splitting of
approximately 2.7 eV (4.0 eV) in the $e_g$ ($t_{2g}$) sub-bands. The
Nb atoms, on the other hand, possess an empty $d$-shell, consistent with the
formal valency of +5. The octahedral ligand field environment also
splits these states into $t_{2g}$ and $e_g$ levels which lie at
approximately 2.6 eV and 7 eV, respectively, corresponding to a ligand field
splitting of approximately 4.5 eV. The much larger CF splitting of the
Nb-4$d$ states compared to Fe-3$d$ can be understood by the larger
spatial extension of the former. However, the exchange splitting is
negligible. The magnetic moment on the Fe atoms is found to $\sim 5 \mu_{\rm
B}$,
consistent with the $d^5$ occupation. 

The ground state within GGA is found to be gapless
(semi-metallic) in sharp contrast to the diffuse reflectance
measurements (UV-Vis-NIR spectra). This is not too surprising for the
applied GGA.
Fig. \ref{fig:dft}(b) shows the GGA band structure. It is interesting to note that
the spin-resolved band structure has an indirect gap for the up-spin
between the high symmetry points L ($\pi/a$, $\pi/a$, $\pi/a$) and X (0, 2$\pi/a$, 0).

The electron-electron correlations beyond GGA are accounted for via GGA$+U$
calculations. Since the Nb-$d$ states are unoccupied, it may suffice to
only apply $U_{\rm eff} = U-J$ (with $J=0$) to the Fe-3$d$ shell. 
With $U$, a gap opens and increases with
increasing values of $U$ (see \ref{sec:AppD} for more details).
Figs. \ref{fig:dft}(c)-(f) show the evolution of DOS with $U$. With
increasing values of $U$, the
unoccupied Fe $d-t_{2g}$ (occupied Fe $d-e_g$) states shift higher (lower) in
energy. The Nb-$d$ states, however, remain unaffected. For the entire range
of considered $U$-values, the ground state corresponds to a half-filled
Fe-$d$ shell, with a net spin moment of approximately $5 \mu_{\rm B}$  per Fe atom as
was the case for $U=0$.

For $U \sim 6$ - $7$ eV, the bandgap is approximately 2 eV in agreement with the
value obtained from the diffuse reflectance measurement. However, it is
important to note that the gap remains indirect for all values of U up
to 7.5 eV and corresponds to a spin-flip electronic transition (possibly
assisted by phonons). The origin of such a discrepancy is likely due to
the fact that the DFT calculations are performed for the $Fm\bar{3}m$ structures
with an ordered occupation of the B-sites by Fe and Nb atoms whereas the
NPD data suggest a ``disordered" arrangement of B-site cations.
Therefore, it is instructive to carry out a detailed investigation of
possible structures with random occupation of B-site cations by
considering larger supercells \cite{Morrow2020}, which is beyond the scope of the present
work.

\begin{figure}
    \centering
    \includegraphics[angle=0,width=0.48\textwidth]{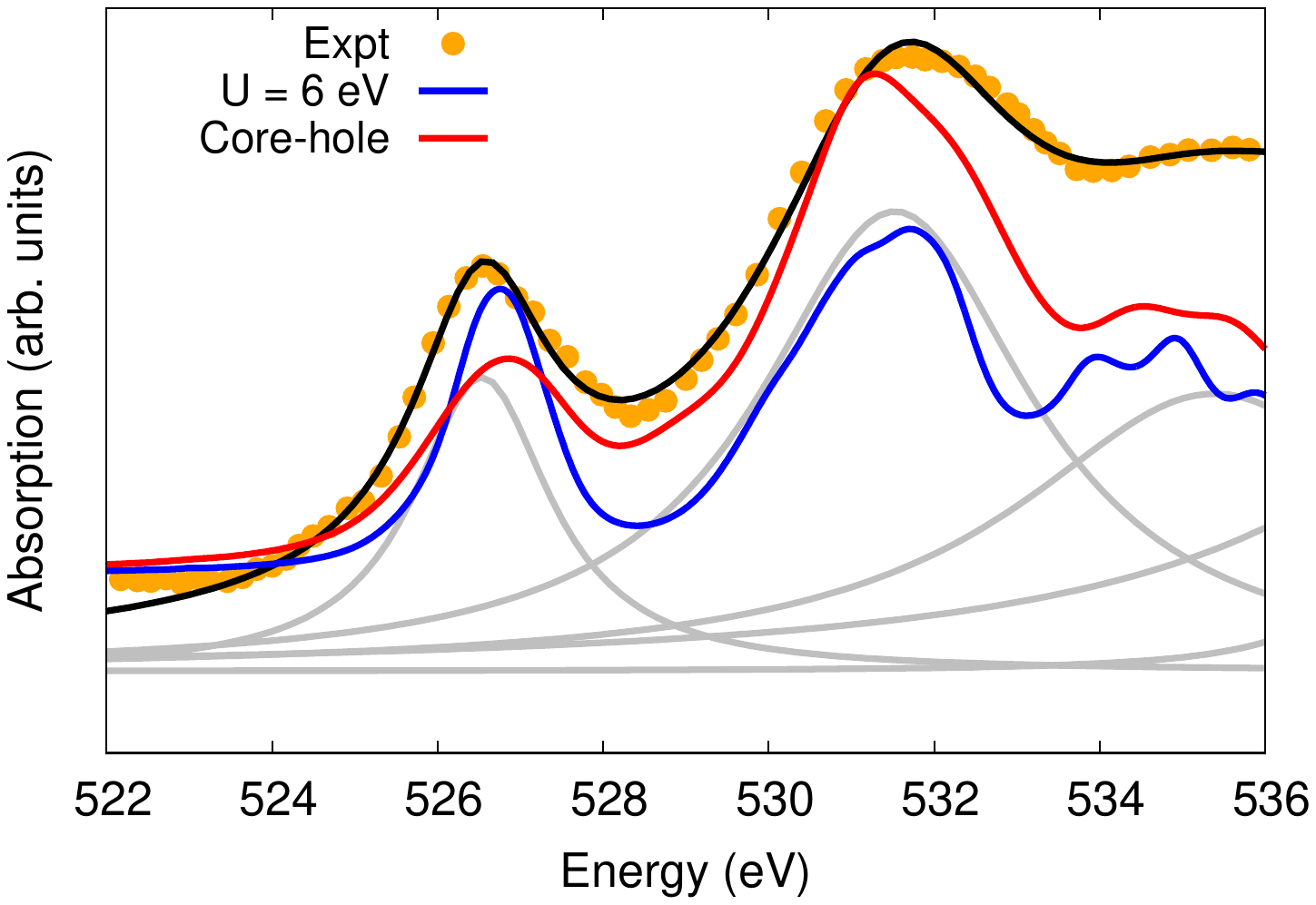}
    \caption{Comparison of the theoretical and experimental XAS O K-edge
        spectra. A fit of the peak positions in the experimental
        spectrum using Lorentzian functions is shown in gray solid
        curves. The resultant curve is shown with black solid line. The
        theoretical spectra was obtained with (red solid curve) and
        without (blue solid curve) considering core hole in the O-1$s$
        state, for $U_{\rm eff} = 6$ eV.
    }
   	\label{fig:ch}
\end{figure}

In order to tune the estimate of $U$ for BFN, we compare the DFT results
with the XAS spectra. Specifically, we compare the position of TM-$d$
states in the conduction band for several values of $U$ with the low-energy
part of the O K-edge spectra. We find that, indeed, the experimental XAS
spectrum is well explained by considering $U = 6$ eV for Fe-3$d$ states.
Fig. \ref{fig:ch}
shows a comparison between the experimental and simulated spectra. The
low-energy peak positions are identified by performing a multiple-peak
(Gaussian) fitting of the experimental curve. The peak positions in the simulated
(theoretical) spectra compare well with the experimental values.
Consideration of core hole in the O-1$s$ state influences the relative
intensity of these low-energy peaks and leads to an improved agreement with the
experimental data. Looking at the partial DOS for the TM-$d$ states (Fig
\ref{fig:dft}), it is clear that the low energy peak at approximately 527 eV is due
to hybridized states from both Fe-$d$ and Nb-$d$ with O-2$p$, whereas the peak
at approximately 532 eV is largely due to hybridization between O-2$p$ and
Nb-$e_g$ states. %It is interesting to note that consideration of core-hole,
%even without the full relaxation for the core-hole, leads to a very good
%qualitative and quantitative  agreement with the experiments. 
With regard to the values of $U$, a similar value for Fe-$d$
states was also suggested for a related compound
Pb(Fe$_{1/2}$Ta$_{1/2}$)O$_3$ \cite{bharti2012}.

\subsubsection{\label{sec:exafs}EXAFS}

\begin{figure*}
    \centering
    \includegraphics[angle=0,width=0.750\textwidth]{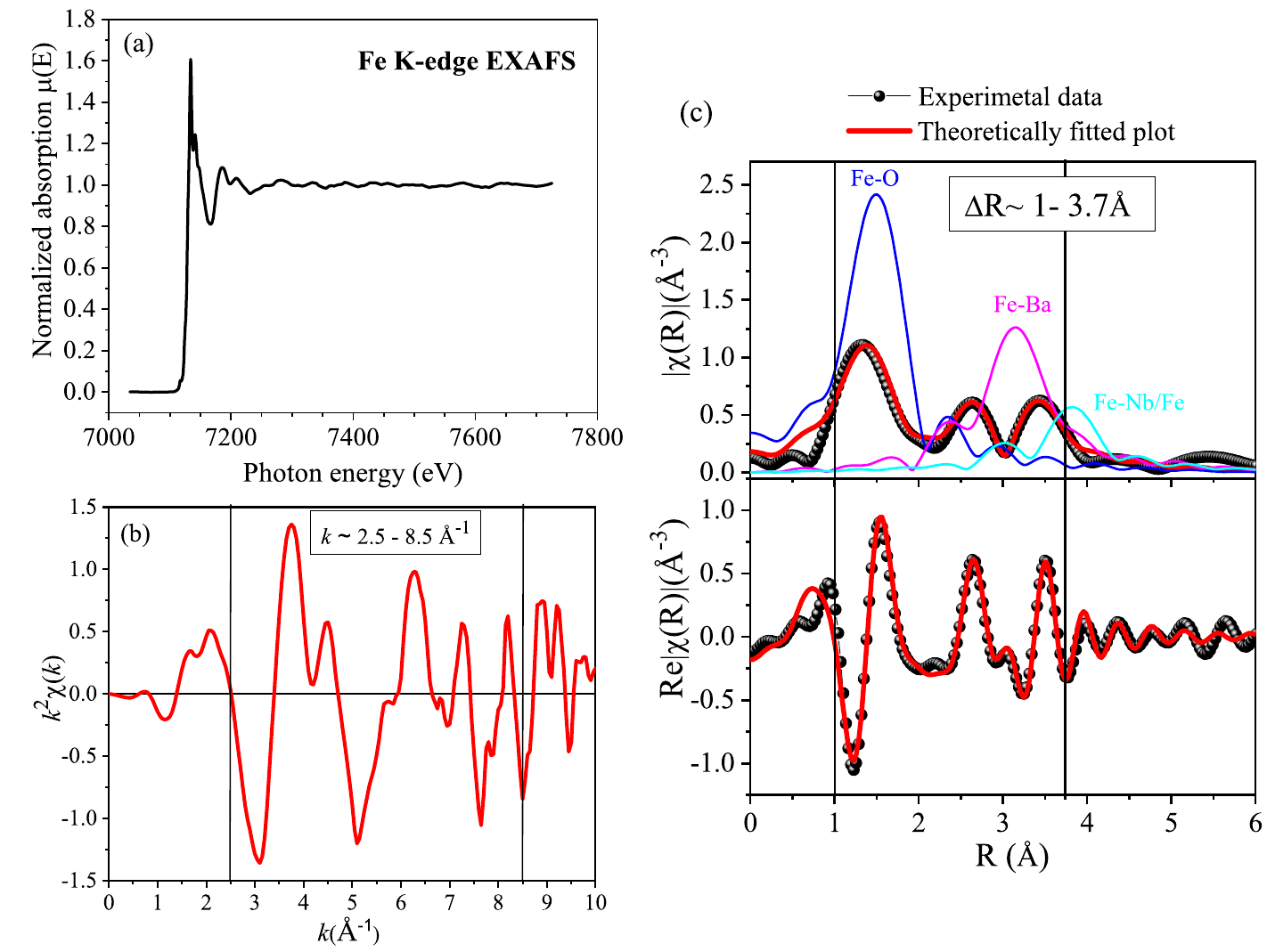}
    \caption{(a) Normalized EXAFS spectrum at the Fe K-edge. (b) $k^2$
        weighted $\chi(k)$ spectra at Fe K-edge, and (c) the
        theoretically generated model (red line) with contributions for
        atomic pairs: Fe-O (blue), Fe-Ba (pink) and Fe-Nb/Fe (cyan).
    }
   	\label{fig:exafs}
\end{figure*}

\begin{table}[t]
    \centering
    \small
    \caption{ EXAFS best fitted parameter values. The bond length values in brackets correspond to
    the structure obtained from the NPD data.}
\label{table:exafs_fit}
\begin{tabular}{lcc}
 
\hline \hline
    & Parameters & BFN \\
\hline \hline
    
    Scattering Path & $S_0^2$ & $0.85$ \\
    $\times$ C.N.  & $E_0$ (eV) & $-5.976 \pm 0.40$ \\
    
    \hline
    Fe$-$O $\times 6 $ & $R$ ({\AA})  & $2.00 \pm 0.01$ \\ 
     (2.023 {\AA})  & $\sigma^2$ ({\AA}$^2$) & $0.014 \pm 0.001$ \\
    
    \hline 
    Fe$-$Ba $\times 8 $ & $R$ ({\AA})  & $3.41 \pm 0.02$ \\ 
     (3.504 {\AA})  & $\sigma^2$ ({\AA}$^2$) & $0.025 \pm 0.004$ \\
    
    \hline 
    Fe$-$Nb/Fe $\times 6 $ & $R$ ({\AA})  & $3.77 \pm 0.01$ \\ 
     (4.047 {\AA})  & $\sigma^2$ ({\AA}$^2$) & $0.003\pm 0.001$ \\
    
    \hline
        &  $R_{\rm factor}$  & 0.012 \\
    \hline
 \end{tabular}
\end{table}

To probe the local structural aspects of the BFN sample, we
carried out EXAFS studies \cite{shuvaeva2017}. Fig. \ref{fig:exafs}(a)
shows the Fe K-edge EXAFS spectra ($\mu(E)$ vs. $E$ data) of the
sample, from which the equivalent absorption function $\chi(k)$ was
obtained. 
Fig. \ref{fig:exafs}(b) shows the $k^2$ weighted $\chi(k)$ vs.
$k$ plot, and the $\chi(R)$ versus $R$ plot is shown in
Fig. \ref{fig:exafs}(c). %along with
%the best fit theoretical simulations. 
In the $\chi(R)$ versus $R$ plot, there are three well-defined
peaks lying between 1 {\AA} and 4 {\AA}. The peak around 1.5 {\AA}
corresponds to the Fe-O bond length. This peak contains
information regarding the off-center displacement of Fe ions and
distortions of the FeO$_6$ octahedra. It is interesting to note that
there is only one well-defined and reasonably sharp peak, confirming the negligible
off-center (non-centrosymmetric) displacement of the Fe
ions \cite{shuvaeva2017}.
The peak around 3.5 {\AA} can be attributed to the interatomic distance
between the B-site cations (Fe-Fe or Fe-Nb distances).

We have also theoretically simulated these features assuming the
structure of the sample obtained from the NPD measurements. 
The comparison between the
simulated and experimental spectra along with the corresponding 
interatomic distances are also shown in Fig \ref{fig:exafs}(c). 
Evidently, the qualitative and the quantitative agreement between the
two curves is very good in the range of 1 $-$ 3.7 {\AA}.
The corresponding best fit results of the fitting parameters ($R$ and $\sigma^2$) are
shown in Table \ref{table:exafs_fit}. 

Our primary interest is in the Fe-O distance since this contains
information regarding the local structure. We find that the
theoretical estimate of 2.0 {\AA} is in a very good agreement with the
experimental XRD/NPD value. The Fe-Ba and Fe-Nb/Fe distances however show
deviations of approximately 0.1 {\AA} and 0.3 {\AA}. The peak
corresponding to the interatomic
distance between the B-site cations in perovskites materials is
sensitive not only to their distance but also to the bond angle
TM-O-TM (TM=Fe/Nb). Deviation
between the theoretical and experimental estimates is likely caused by the
relative difference in electronegativity and oxidation state of the two
ions and is indicative of the fact that the BO$_6$ and
B$'$O$_6$ octahedra are slightly distorted. Nevertheless, the average Fe-O
distances turn out to be in good agreement with the expected value,
implying that these distortions are non-polar in nature.
Therefore, the local structure, especially the distances between transition metals as well as
    with oxygen, obtained via EXAFS is in an excellent agreement with the crystal structure
    determined by NPD.

\section{Conclusions}
\label{sec:concl}
In summary, we have carried out an in-depth study of the structural and
electronic properties of the BFN ceramics. %which show high dielectric constant. 
Combined XRD, NPD and EXAFS studies put to rest the existing
discrepancies related to the structural details of this material. The
study reveals a cubic $Pm\bar{3}m$ structure ($a = 4.0556$ {\AA} at 300
K) with the chemical formula corresponding to the perovskite {\bfn}. 

More importantly, the presented analysis of the available structural reports
highlights subtleties in structural characterization of complex
perovskites with high pseudosymmetry.
In such cases, it may be particularly challenging to correctly characterize the structural
symmetry and, therefore, complementary techniques should be
employed. Combined XRD and NPD measurements together with a careful
analysis of the data provide such a framework.

Regarding the
electronic properties, a good agreement is found between XAS and
DFT studies. The formal valency of B-site transition metals Fe and Nb
is found to be +3 and +5, respectively. Consequently, the Fe-3$d$ is
half-filled with spin up electrons while the Nb-4$d$ states are unfilled,
leading to a net spin moment of 5 $\mu_{\rm B}$ on each Fe atom.
The ground state is insulating. The gap is estimated from the diffuse reflectance
measurement in the UV-Vis-NIR range and the estimated gap of
approximately 2.0 eV is consistent with the sample color. 

From a theoretical standpoint, DFT calculations for an $Fm\bar{3}m$
structure model (considering a $2 \times 2 \times 2$ supercell) within
GGA provides a good qualitative
and quantitative description of the electronic properties. A rather
large value of $U_{\rm eff} \sim 6$ eV is needed to account for the
correlation effects. However, it turns out that the nature of 
bandgap is inconsistent with the experimental results and larger
supercells must be considered to account for the random B-site
occupation and/or distortion in the BO$_6$ octahedra.

\vspace{1cm}
\textit{Note added: During the preparation of the manuscript, we became aware of a recent study
    which also reported a cubic $Pm\bar{3}m$ structure using high resolution synchrotron X-ray powder
diffraction (SXRPD) measurement. \cite{Kumar2021}}

\section*{Additional Information}
\textbf{Competing interests.} The authors have no conflicts to disclose.

\section*{Acknowledgments}
We thank Dr. Kaustava Bhattacharya, Dr. Steffen Oswald, Dr. B. K. Singh, and Dr. Sai Aswartham for helpful discussions. RR thanks Ulrike
Nitzsche, Valligatla Sreeramulu and V.V. Aneesh for technical support.
RR and MR acknowledge financial support from
the European Union (ERDF) and the Free State of Saxony via the ESF
project 100339533 (Young Investigators Group Computer Simulations for
Materials Design $-$CoSiMa) during the initial part of the project. 
%\end{section*}

\section*{Data availability statement}
The data that support the findings of this study are available from the corresponding author upon
reasonable request.

\flushbottom
\newpage
\newpage
    
%\appendix
\begin{appendix}

%\begin{widetext}

\section{Structural characterization of \texorpdfstring{FeNbO$_4$}{fenbo4}}
\label{sec:AppA}

\begin{figure}[hb!]
    \centering
    \includegraphics[angle=0,width=0.48\textwidth]{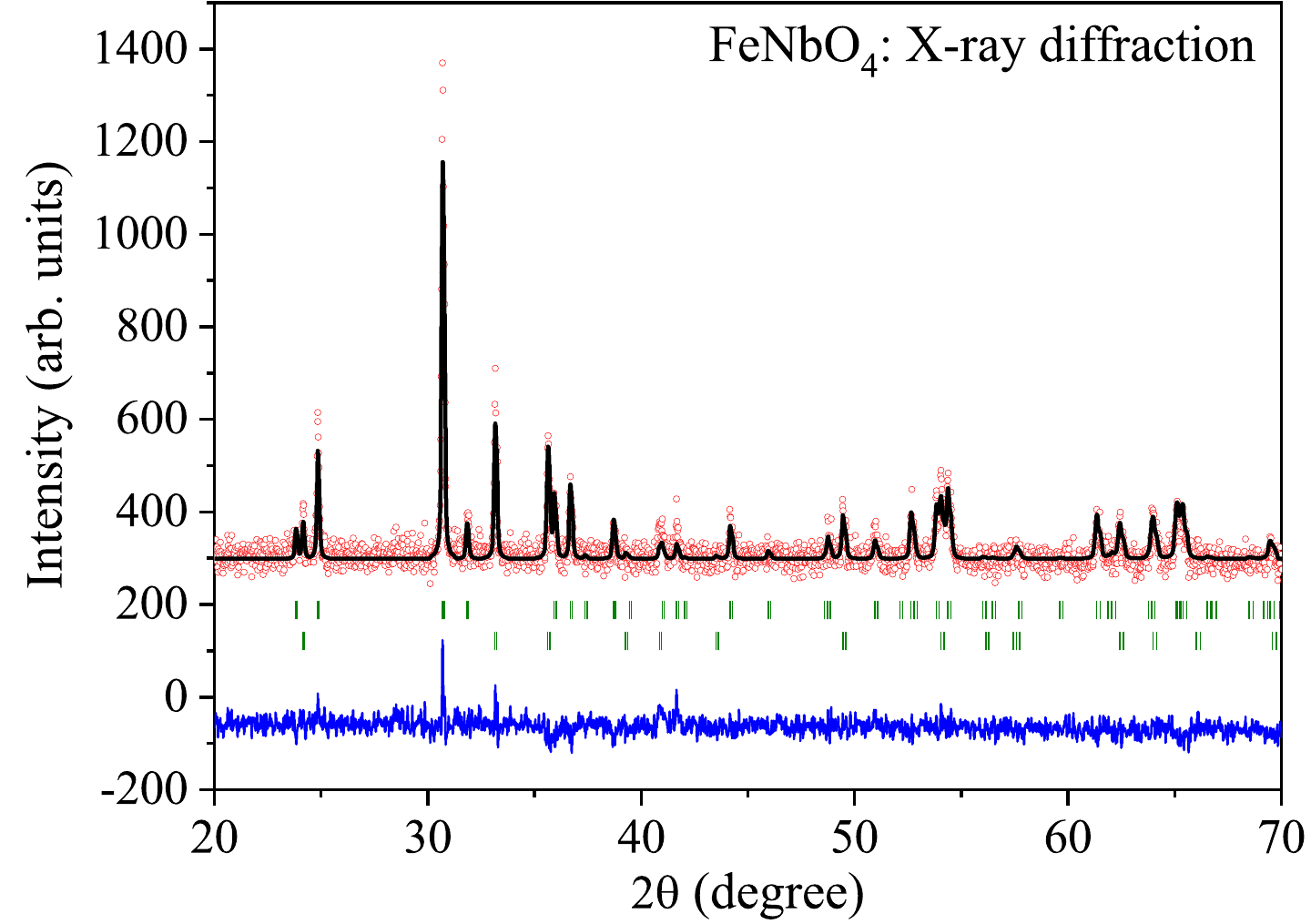}
    \caption{XRD pattern and Rietveld refinement of the precursor. The Bragg positions are
        shown with the vertical bars for both FeNbO$_4$ (top) and
        Fe$_2$O$_3$ (bottom). The crystal structure of FeNbO$_4$ is
        identified to be the monoclinic $P2/c$ structure.
    }
    \label{fig:xrd_fno}
\end{figure}

Figure \ref{fig:xrd_fno} shows the XRD pattern of the precursor
composed of FeNbO$_4$ and Fe$_2$O$_3$. The corresponding Rietveld
refinement suggests a monoclinic structure for FeNbO$_4$. 

%\section{\label{sec:AppB} NPD data for the \texorpdfstring{$Fm{\bar 3}m$}{fm3m} structural model}
\section{\label{sec:AppB} Analysis of the NPD data for other structural models}

\begin{figure}[ht!]
    \centering
    \includegraphics[angle=0,scale=0.325]{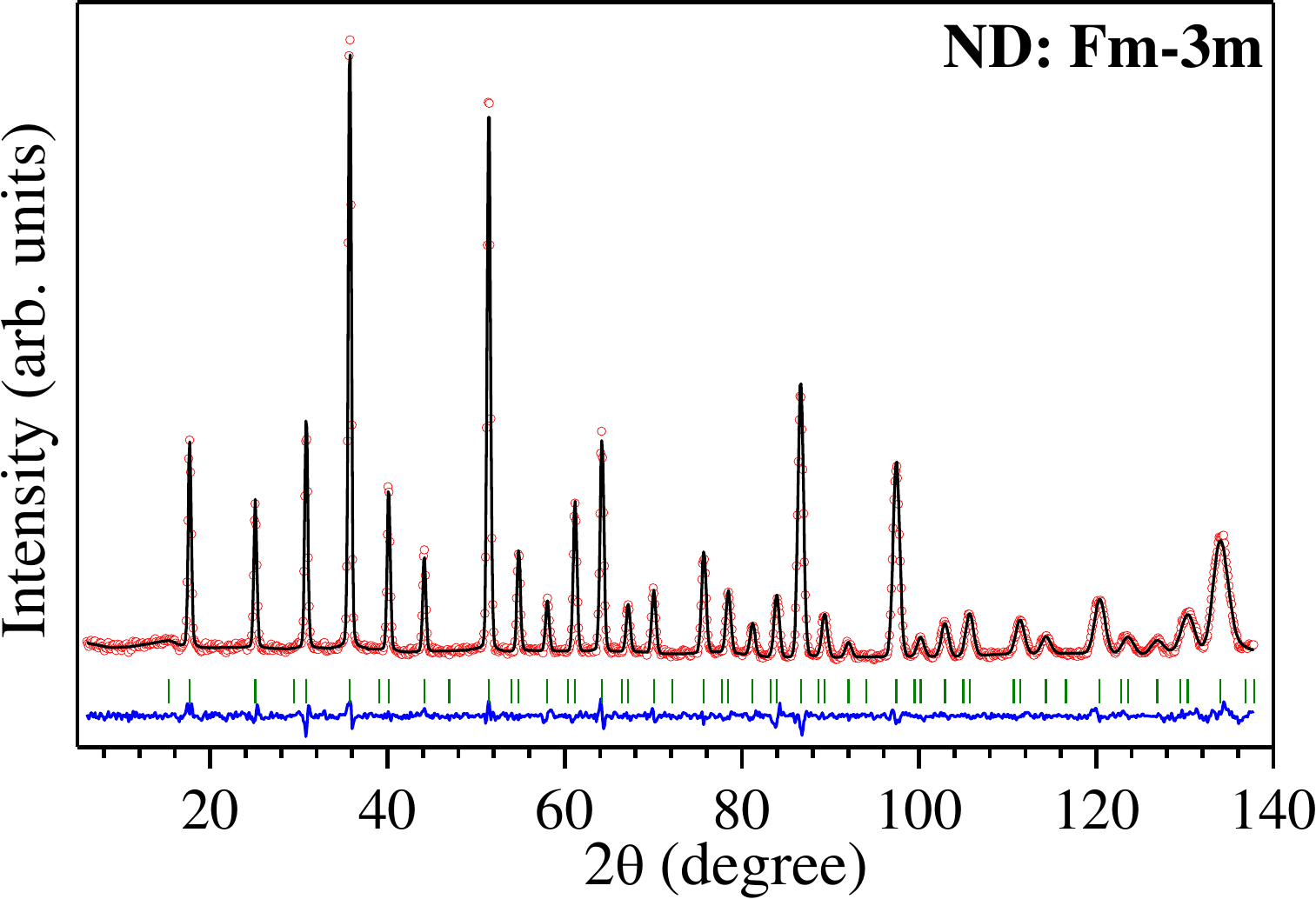}
    \caption{ Fit of the room temperature NPD data to the $Fm{\bar 3}m$ structural model.
    }
   	\label{fig:npd_fm3m}
\end{figure}

Figure \ref{fig:npd_fm3m} shows the fit of the $Fm{\bar 3}m$ structural model to the NPD
data. %, as compared to the $Pm{\bar 3}m$ model.
The final R-indices for the fit to the $Fm{\bar 3}m$ structure are: $R_{p} = 3.94$,  $R_{wp} = 5.02$, and  $R_{exp} = 3.70$, and 
the GoF = 1.92. Therefore, the quality of the fit is comparable to the $Pm{\bar 3}m$
structural model (see Table \ref{table:str_npd}). However, despite the quality of the fit, any sublattice reflection peaks corresponding to deviation of the oxygen
position from the center of the Fe/Nb positions are absent, ruling this out as an appropriate
structural model for the synthesized BFN ceramics.

\begin{figure}[ht!]
    \centering
    \includegraphics[angle=0,scale=0.350]{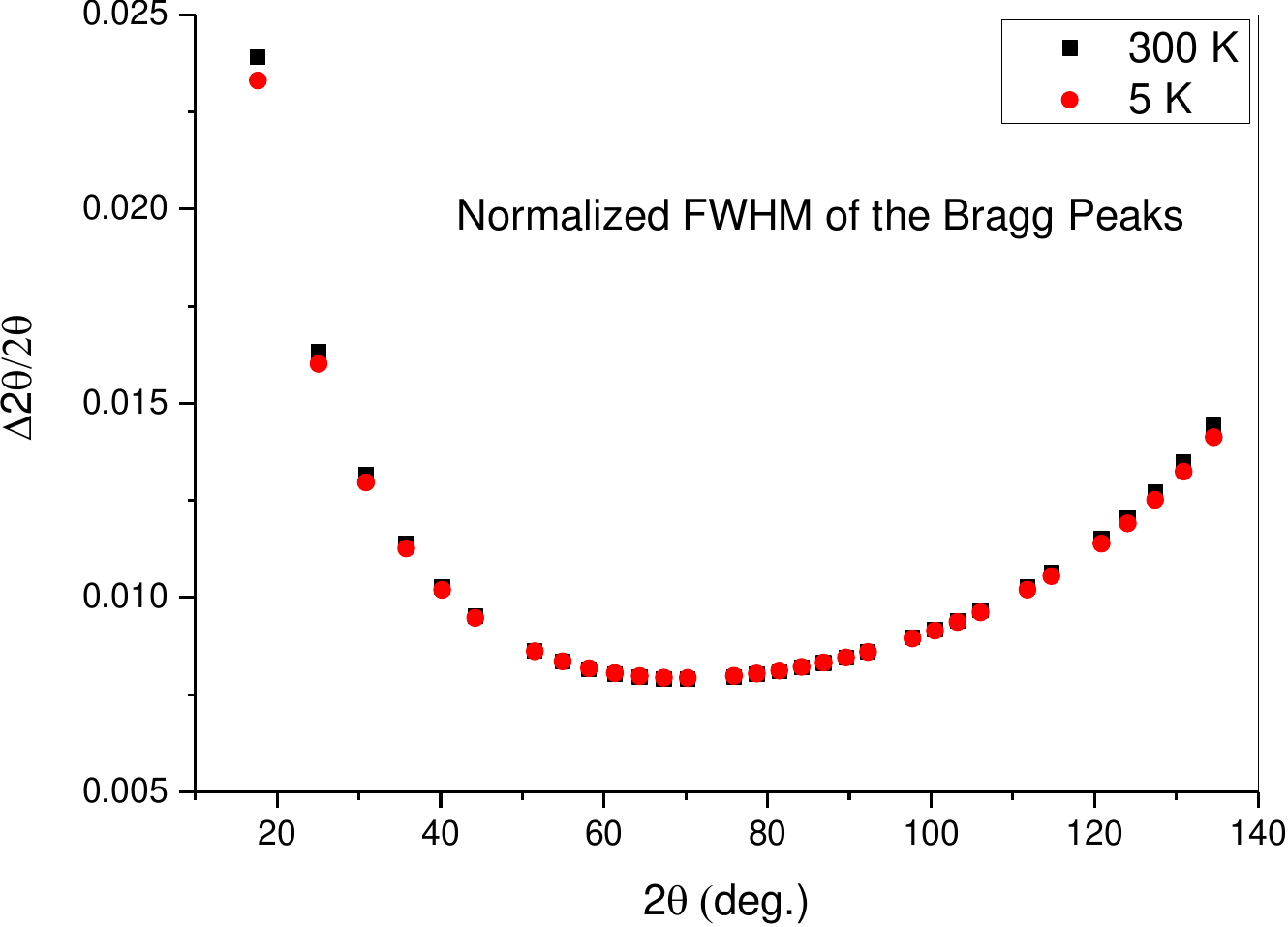}
    \caption{ Normalized FWHM of the Bragg peaks of the NPD data at $300\,$K and $5\,$K, showing absence of
        structural distortion.
    }
   	\label{fig:npd_fwhm}
\end{figure}

Similarly, no superlattice reflection peaks were observed for the monoclinic structure.
    To investigate any subtle structural transformation at low temperature, we study the
    full-width half-maxima (FWHM) of the Bragg peaks in the NPD data as well. Figure
    \ref{fig:npd_fwhm} shows that
    the FWHM does not change between $300\,$K and $5\,$K, thus ruling out structural transitions.

\section{\label{sec:AppC}Considered structures for DFT calculations}

Table \ref{table:supercell_str} presents the structural characteristics
and relative spin-polarized GGA energies of various possible structures
in a ($2 \times 2 \times 2$) supercell. The supercell structure contains
four Fe atoms and four Nb atoms. Due to the size of the supercell, all
    the resulting structures consist of a periodic arrangement of the TM
    ions:
    \begin{itemize}
        \item [-] {\bf S1}: alternating layers of Nb and Fe atoms
            along a principle direction,
        \item [-] {\bf S2}: alternating Nb and Fe atoms along all
            principle directions (leading to a face centered cubic
            structure),
        \item [-] {\bf S3}: alternating Nb and Fe atoms along a face
            diagonal.
        \item [-] {\bf S4}: alternating Nb and Fe atoms along a face
            diagonal and the principle direction orthonormal to it.
    \end{itemize}
%            . The structures , {\bf S2} and {\bf S3}, respectively, correspond to cases where 
%the Nb atoms lie in a plane, alternatively with Fe 
%atoms along one principle direction, and along all principle
%    directions, 
%and along one face diagonal and orthonormal principle direction.
Considering the
$Pm\bar{3}m$ structure obtained by XRD and NPD data, the position of
the oxygen atom was kept fixed. For example, for the cubic $Fm\bar{3}m$
structure ({\bf S3}), the position of O atoms were fixed at (1/4,0,0).

\begin{table}[h!]
    \centering
    \small
    \caption{Relative energies of various considered structures for
    {\bfn} obtained by considering supercells. The corresponding lattice
    parameters are also listed with respect to the lattice parameter
    obtained from NPD refinement $a = 4.0556$ {\AA}.
    }
\label{table:supercell_str}
\begin{tabular}{lccc}
 \hline 
    \hline
    {\bf Case} & {\bf Space group} & {\bf Lattice params} & {\bf Energy/f.u. (eV)} \\
 \hline \hline
    {\bf S1}   &   $P4/mmm$    &  ($a$,$a$,$2a$)    &   0.223 \\
    {\bf S2}   &   $Fm\bar{3}m$  &  ($2a$,$2a$,$2a$)    & 0.0 \\
    {\bf S3}   &   $P4/mmm$  &  ($\sqrt{2}a$,$\sqrt{2}a$,$a$)    &   0.071 \\
    {\bf S4}   &   $P4/mmm$  &  ($2a$,$2a$,$2a$)    &   0.088 \\
    \hline
    \hline
\end{tabular}
\end{table}

\begin{figure}[h!]
    \centering
    \centerline{\includegraphics[angle=0,width=0.495\textwidth]{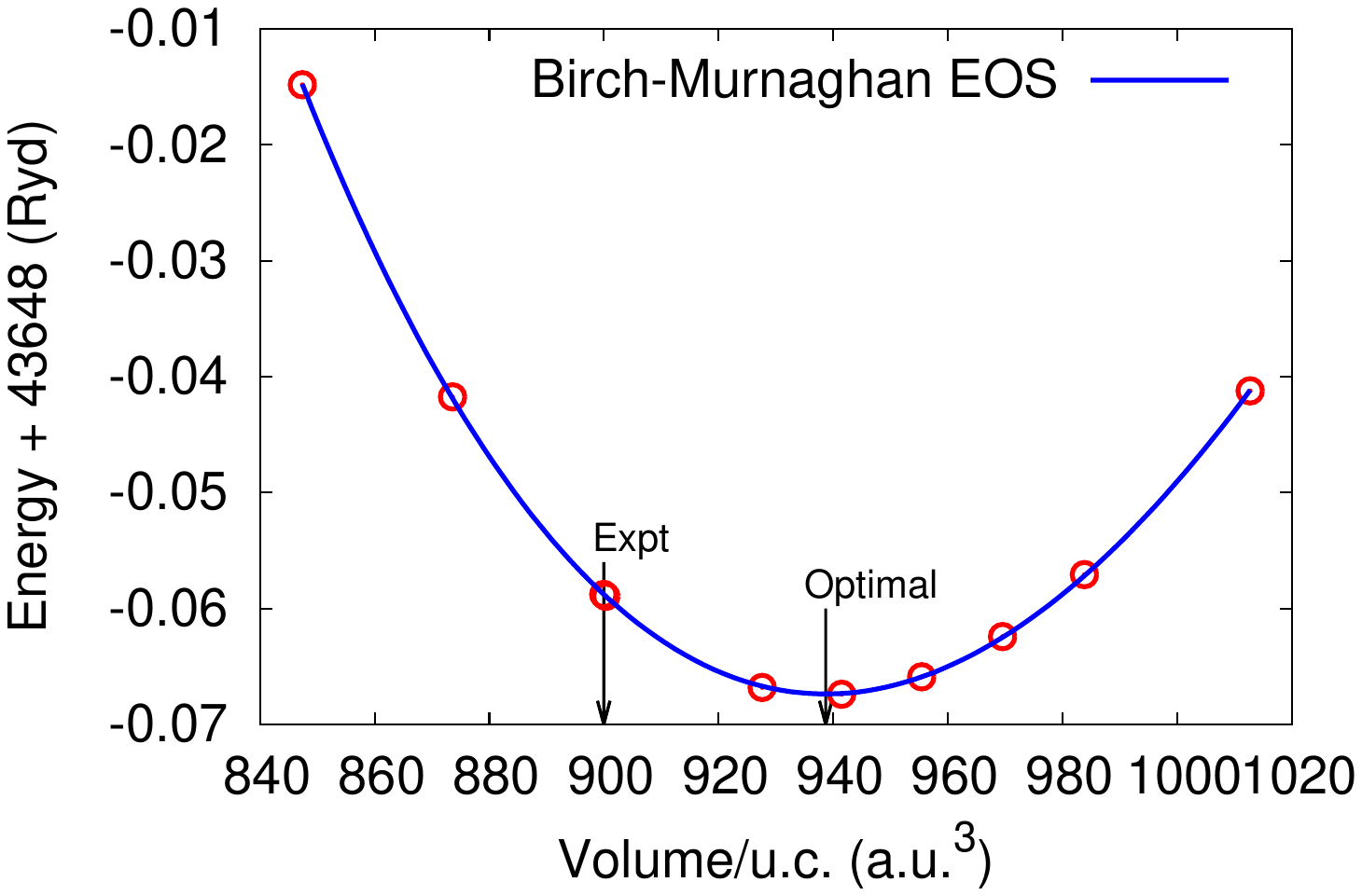}}
    \caption{
        Energy vs volume for the $Fm\bar{3}m$ representation of BFN, with oxygen position fixed. The open
       circles are the different considered volumes while the solid
       curve is the fitted Birch-Murnaghan equation of state (EOS). The optimal
       unit cell volume is 938.66 a.u.$^3$, which is approximately 4.3\%
       larger than the experimental value. Correspondingly the lattice
       constants are approximately 1.4\% larger.}
   	\label{fig:vol-opt}
\end{figure}

The face-centered cubic
$Fm\bar{3}m$ structure turns out to have the lowest energy and was used
for a detailed analysis of the electronic properties. We also obtained
the optimal lattice parameters for the $Fm\bar{3}m$ structure. Fig.
\ref{fig:vol-opt} shows the dependence of energy on volume of the unit
cell, as obtained within the spin-polarized GGA calculations and
accounting for scalar relativistic corrections.

\section{\label{sec:AppD}Evolution of gap with \texorpdfstring{$U$}{U}}

Table \ref{table:gap_u} shows the evolution of the spin-resolved and net bandgap with $U$.

\begin{table}[h!]
\centering
\small
    \caption{Electronic bandgap as a function of $U$. 
    }
\label{table:gap_u}
\begin{tabular}{cccc}
 \hline 
    \hline
    {\bf $U$-value} (eV) & \multicolumn{3}{c}{\bf Gap (eV)} \\
                   & $\uparrow$-spin & $\downarrow$-spin  & Net  \\
 \hline \hline
    0   &   1.07    &   0       &   0 \\
    1.0 &   2.38    &   1.27    &   0.40 \\
    3.0 &   2.75    &   1.58    &   1.09 \\
    5.0 &   2.99    &   1.87    &   1.65 \\
    6.0 &   3.10    &   2.01    &   1.89 \\
    6.5 &   3.13    &   2.07    &   1.99 \\
    7.0 &   3.15    &   2.13    &   2.10 \\
    7.5 &   3.18    &   2.19    &   2.19 \\
    \hline
    \hline
\end{tabular}
\end{table}
%\vspace{3.5cm}

\clearpage
\newpage

\begin{widetext}

%\section[textwidth=2\textwidth]{Review of reported structures and synthesis parameters}
\section{\label{sec:AppE}Review of reported structures and synthesis parameters}

%\begin{widetext}

%\small{
\begin{longtable}{p{0.18\textwidth} p{0.190\textwidth} p{0.220\textwidth} p{0.15\textwidth} p{0.20\textwidth} }
\caption{Review of reported crystal structures for BFN ceramics
along with the method and synthesis parameters. $T_{\rm cal}$ and
$T_{\rm sin}$, respectively, represent the calcination and sintering
temperatures. The values in brackets are the duration; empty
brackets indicate that the values are not known. }  \\
\hline 
\hline
{\bf Method} & \multicolumn{2}{c}{\bf Synthesis parameters} & {\bf Crystal symmetry} & {\bf Source} \\
& $T_{\rm cal}$ [$^{\circ} C$] (duration) & $T_{\rm sin}$ [$^{\circ} C$] (duration) &  & \\
\hline
\hline
    \\
    %\multicolumn{4}{l}{\textbf{A. Solid state method (without precursor)}} \\
    %\hline
    Solid state method & 900 (2h) & 1350 (4h) & cubic $Pm{\bar 3}m$    & Yokosuka {\it et al.} (1995) \cite{yokosuka1995} \\
    (without precursor) & 900 (2h)  & 1200 (60h) $\rightarrow$ 1300 (12h) & cubic $Fm{\bar 3}m$ & Tezuka {\it et al.} (2000) \cite{tezuka2000} \\ 
    &                  1200 (10h) & 1250 (5h) & monoclinic  & Saha \& Sinha (2002)\cite{saha2001}\\
    &                  997 - 1097 (4h) & 1197 - 1547 () & cubic $Pm{\bar 3}m ^{\dagger}$ & Raevski {\it et al.} (2003) \cite{raevski2003} \\ 
    &                  1100 (4h) & 1300 () & cubic $Fm{\bar 3}m$ & Shuvaeva {\it et al.} (2003) \cite{shuvaeva2017} \\ 
    &                  1200 (24h)  & 1350 ()  & cubic $Fm{\bar 3}m$    & Rama {\it et al.} (2004) \cite{rama2004} \\
    &                  997 - 1097 (4h) & 1150 ()\newline 1280 - 1350 () & cubic $Fm{\bar 3}m ^{\dagger}$ \newline monoclinic    & Demirbilek {\it et al.} (2004) \cite{demirbilek2004} \\
    &                  800 - 1200 (4h) & 1250 - 1400 (4h)  & cubic $Fm{\bar 3}m$ & Eitssayeam {\it et al.} (2006) \cite{eitssayeam2006} \\ 
    &                  800 - 1200 (4h) & 1250 - 1400 (4h)  & cubic $Fm{\bar 3}m^{\ddagger}$ & Eitssayeam {\it et al.} (2009) \cite{eitssayeam2009} \\ 
    &                  1200 (3h) & 1350 (3h) & cubic $Pm{\bar 3}m$    & Wang {\it et al.}(2007) \cite{wang2007} \\ 
    &                  1200 (5h) & 1250 (4h) & cubic $Pm{\bar 3}m$    & Bhagat \& Prasad (2010) \cite{bhagat2010} \\ 
    &                  1200 (4h) & 1350 (4h) & cubic $Pm{\bar 3}m$    & Intatha {\it et al.} (2010) \cite{intatha2010} \\ 
    &                  1000 (2h) $\rightarrow$ 1200 (2h) & 1250 (4h) & cubic $Pm{\bar 3}m$  & Ganguly {\it et al.} (2011) \cite{ganguly2011} \\ 
                     % & 1000 (4h) & 1200 - 1400 () & cubic $Pm{ar 3}m$,\newline monoclinic & Raevski {\it et al.} (2009) \cite{raevski2009b} \\ 
    &                 900 - 1200 (4h) & 1000 - 1200 (2h) & cubic $Pm\bar{3}m$ & Khopkar \& Sahoo (2020) \cite{Khopkar2020} \\
    \\
    %\multicolumn{4}{l}{\textbf{B. Columbite precursor}} \\
    %\hline
    Columbite precursor &  950 (5h) & 1200-1350 (4h) & monoclinic & Chung {\it et al.} (2004) \cite{chung2004} \\
     & 1200 (8 h) & 1250, 1300, 1350 (6h) & monoclinic  & Ke {\it et al.} (2008) \cite{ke2008}\\
     & 950 (5h) & 1200-1300 (4h) & monoclinic & Chung {\it et al.} (2008) \cite{chung2008} \\
     & 1200 (8h) & 1300 (6h)  & cubic $Fm{\bar 3}m$ & Ke {\it et al.} (2009, 2013) \cite{ke2009, ke2013} \\
     & 1000 (8h) & 1100 (8h)  & cubic $Pm{\bar 3}m$    & Liao {\it et al.} (2010) \cite{liao2010} \\
     & 1200 (8h) & 1300 (6h)  & cubic $Pm{\bar 3}m$    & Ke {\it et al.} (2010) \cite{ke2010} \\
     & 1200 (8h) & 1250 (4h) & cubic $Pm{\bar 3}m$    & This work \\
    \\
    %\multicolumn{4}{l}{\textbf{C. Traditional mixed oxide}} \\
    %\hline
    Traditional mixed oxide & 1200 ()  & 1250,1300,1350 ($\gtrsim$4h)$^*$ & monoclinic & Ke {\it et al.} (2008) \cite{ke2008}\\
    \\
    %\multicolumn{4}{l}{\textbf{D. Powder calcination }} \\
    %\hline
    Powder calcination & 1250 (4h) & 1350 (2-5h) & monoclinic & Bochenek {\it et al.} (2009) \cite{bochenek2009_bfn}\\
    \\
    %^\multicolumn{4}{l}{\textbf{E. Chemical}} \\
    %\hline
    Chemical & 1300 (12h) & 1400 (4h) & monoclinic & Charoenthai {\it et al.} (2008) \cite{charoenthai2008}\\
    \\
    %\multicolumn{4}{l}{\textbf{F. Sol-gel}} \\
    %\hline                     
     Sol-gel & 550 - 850 (5h) & 1100 -1250 (3h) & monoclinic & Chung {\it et al.} (2005) \cite{chung2005}\\
    \\
    %\multicolumn{4}{l}{\textbf{G. Mechanical triggering}} \\
    %\hline  
    Mechanical triggering & $-$  & $-$ & monoclinic & Bochenek {\it et al.} (2015) \cite{bochenek2015}\\
    \\
    %\multicolumn{4}{l}{\textbf{H. Mechanochemical}} %\\
    %\hline
     Mechanochemical & $-$  & $-$ & cubic & Bochenek {\it et al.} (2018) \cite{bochenek2018}\\
    \hline
    %\hline
    \multicolumn{4}{l}{$^\dagger$ estimated from the discussion. Lattice parameters are not available.} \\
    \multicolumn{4}{l}{$^\ddagger$ estimated from the lattice parameters provided.} \\
    \multicolumn{4}{l}{$^*$ estimated value from a cooling rate of 5$^{\circ}$C/min.} \\
\label{table:review}
\end{longtable}
%\end{table*}
%}

\end{widetext}

\end{appendix}

%\clearpage
%\newpage
%\section*{References}
%\nocite{*}
%\bibliography{bfn}% Produces the bibliography via BibTeX.
%

\end{document}